\documentclass[conference]{IEEEtran}
\IEEEoverridecommandlockouts
\usepackage{cite}
\usepackage{amsmath,amssymb,amsfonts}
\usepackage{algorithmic}
\usepackage{url}
\usepackage{graphicx}
\usepackage{textcomp}
\usepackage{xcolor}
\usepackage{subfigure}
\def\BibTeX{{\rm B\kern-.05em{\sc i\kern-.025em b}\kern-.08em
    T\kern-.1667em\lower.7ex\hbox{E}\kern-.125emX}}
\begin{document}

\newcommand{\othertm}{\textsuperscript{$\star$}}
\newcommand{\regtm}{\textsuperscript{\textregistered{}}}
\newcommand{\tm}{{\scriptsize\texttrademark{}}}
\newcommand{\code}[1]{\begin{small}{\texttt{#1}}\end{small}}

\title{Intel\regtm{} SHMEM: GPU-initiated OpenSHMEM using SYCL\\
}

\author{\IEEEauthorblockN{Alex Brooks, Philip Marshall, David Ozog, Md. Wasi-ur- Rahman, Lawrence Stewart, Rithwik Tom \textsuperscript{\textsection}}
\IEEEauthorblockA{\textit{Intel Corporation} \\
\{alex.brooks, philip.marshall, david.m.ozog, md.rahman, lawrence.stewart, rithwik.tom\}@intel.com}
}

%\and
%\and
%\and
%\and
%\and

\maketitle

\begingroup\renewcommand\thefootnote{\textsection}
\footnotetext{All authors contributed equally}
\endgroup

\let\thefootnote\relax\footnote{\hspace{-3mm}© 2024 IEEE.  Personal use of this material is permitted.
Permission from IEEE must be obtained for all other uses, in any current or
future media, including reprinting/republishing this material for advertising
or promotional purposes, creating new collective works, for resale or
redistribution to servers or lists, or reuse of any copyrighted component of
this work in other works.}

\begin{abstract}
Modern high-end systems are increasingly becoming heterogeneous, providing users 
options to use general purpose Graphics Processing Units (GPU) and other accelerators for 
additional performance. High Performance Computing (HPC) and Artificial Intelligence (AI) 
applications are often carefully arranged to overlap communications and computation for 
increased efficiency on such platforms. This has led 
to efforts to extend popular communication libraries to 
support GPU awareness and more recently, GPU-initiated operations. In this paper, we 
present Intel SHMEM, a library that enables users to write programs that are GPU 
aware, in that API calls support GPU memory, and also support GPU-initiated 
communication operations by embedding OpenSHMEM style calls within GPU kernels. We 
also propose thread-collaborative extensions to the OpenSHMEM standard
that can enable users to better exploit the strengths of GPUs. Our implementation adapts to choose between direct load/store from GPU and 
the GPU copy engine based transfer to optimize performance on different configurations. 

\end{abstract}


\section{Introduction}
\label{sec:intro}

OpenSHMEM~\cite{openshmem}, an open standard first released in 2010, builds on
a legacy of distributed "SHared MEMory" (SHMEM) programming, originally
introduced in 1993 by Cray Research to transfer data across the Cray T3D
machines~\cite{cray-t3d}.
OpenSHMEM specifies a portable library-based API for Partitioned Global Address
Space (PGAS) programming, which enables low-latency one-sided reads
 and writes (\textit{get} and \textit{put}), \textit{atomic} operations, and
\textit{collective} operations that act on remotely accessible data objects.
Version 1.5 of the OpenSHMEM API enables transferring application data and
synchronizing across any, all, or a subset (i.e., a \textit{team}) of the
application's distributed processing elements (PEs)~\cite{teams}.

While OpenSHMEM has proven itself as an efficient and scalable programming
model for CPU-driven HPC; at the time of this writing, OpenSHMEM does not yet
include any standardized support for GPUs.
However, according to the June 2024 Top500 list~\cite{top500}, nine of the top ten
HPC systems include GPU accelerators, that are manufactured by various vendors
like Advanced Micro Devices (AMD), Intel Corporation, and NVIDIA.
The total number of GPUs within the top three systems alone surpasses 115,000:
Frontier has 37,608 GPUs (4 per node), Aurora has 63,744 (6 per node), and
Eagle has 14,400 (8 per node).
Furthermore, each GPU device on both the Aurora and Eagle systems actually
contains 2 graphics compute tiles (or dies), so the total number of
independently usable GPU compute
tiles in the top 3 systems approaches 200,000.
These advanced and disparate architectures necessitate changes in both
OpenSHMEM applications and libraries to support and optimize the execution path
involving GPUs.

Since 2020, NVIDIA has supported NVSHMEM~\cite{b2,b1}, which enables
OpenSHMEM-like APIs to be called within CUDA kernels to support data transfer
across clusters of NVIDIA GPUs~\cite{cuda}.
CUDA provides a general purpose parallel computing platform and programming
model that exploits the compute engines in NVIDIA GPUs to execute user
applications and libraries.
Recently, AMD has announced ROCm OpenSHMEM
(\texttt{ROC\_SHMEM})~\cite{b3,rocshmem} for AMD GPUs leveraging a programming
model similar to CUDA, known as the Heterogeneous-computing Interface for
Portability (HIP)~\cite{hip}.
These solutions have been utilized for several state-of-the-art computational
science use-cases; including PETSc's star-forest graph communication
layer~\cite{petscsf}, QUDA's lattice quantum chromodynamics~\cite{quda},
SV-Sim's quantum computer circuit simulation~\cite{svsim}, cuFFTMp's
large-scale fast Fourier Transforms~\cite{cufft}, and Kokkos Remote
Spaces~\cite{kokkos-remote-spaces}.

In this paper, we present Intel SHMEM, which supports Intel GPU programming
in the System
wide Compute Language (SYCL) environment~\cite{sycl}.
SYCL is a C++ based portable programming model that enables hardware accelerators from
different vendors. The NVSHMEM and ROC\_SHMEM solutions similarly
enable SHMEM operations on their respective vendor GPUs, but at
present they restrict users to their vendor-specific programming
environments.

It seems clear that GPUs should support the existing OpenSHMEM APIs,
but at present, the three efforts offer slightly divergent extension
APIs for the particular needs of massively parallel GPUs. It is not
immediately clear which of the extension concepts are inter-operable.
We feel that Intel SHMEM, designed to be consistent with a portable
programming environment, makes a useful contribution to the expected
standardization effort for GPU extensions to OpenSHMEM.

\iffalse
While the NVSHMEM and ROC\_SHMEM solutions enable SHMEM on their respective
GPUs, they restrict users to their vendor-specific programming
environments.Specifically, there are subtle but important architectural and syntactical
differences between these programming environments.
As examples, the GPU memory and kernel management APIs differ in their names
and often their semantics, NVIDIA hardware executes 32-thread \textit{warps}
while AMD hardware may execute 32 or 64-thread \textit{wavefronts}, and CUDA
decomposes kernels into \textit{blocks} of \textit{grids} while HIP does more
so into \textit{work-groups} of \textit{work-items}.
It is not immediately clear which of these concepts are inter-operable, so
these differences inhibit the portability, and thus standardization, of
OpenSHMEM interfaces that might support GPUs.

In this paper, we present Intel SHMEM, which supports Intel GPUs in a System
wide Compute Language (SYCL) environment~\cite{sycl}.
SYCL is a C++ based programming model that enables hardware accelerators from
different vendors, which greatly improves the GPU programming productivity and
portability.
\fi
Intel SHMEM is a C++ software library that enables applications to use
OpenSHMEM communication APIs with device kernels implemented in SYCL.
At the time of this writing, the implementation specifically supports only the
Intel\regtm{} Data Center GPU Max Series.
However, because SYCL is a cross-platform abstraction layer supporting multiple
vendors' GPUs~\cite{sycl-interop}, the design of Intel SHMEM is well-positioned
to target cross-vendor GPU execution in the near future.

Intel SHMEM implements a Partitioned Global Address Space (PGAS) programming
model and includes most host-initiated operations in the current OpenSHMEM
standard.
It also includes new device-initiated operations callable directly from GPU
kernels. The core set of features that Intel SHMEM provides are:

\begin{itemize}
\item{Device and host API support for OpenSHMEM 1.5 compliant point-to-point Remote 
Memory Access (RMA), Atomic Memory Operations (AMO), signaling, memory ordering, and 
synchronization operations}
\item{Device and host API support for collective operations aligned with the
OpenSHMEM 1.5 teams API}
\item{Device API support for SYCL work-group and sub-group level extensions for RMA, 
signaling, collective, memory ordering, and synchronization operations}
\item{A complete set of C++ function templates that supersede the C11 Generic
routines in the current OpenSHMEM specification}
\end{itemize}

In the sections below, we describe implementation details and trade-offs
regarding these features and how they optimize for runtime performance.
The design and implementation of Intel SHMEM are described in the
context of the Aurora supercomputer~\cite{aurora} architecture including changes needed 
for efficient
interactions between CPU and GPU components.
We finally present performance evaluations using RMA and collective operation
micro-benchmarks that quantify throughput and latency versus critical
parameters such as the message size, the work-group size, and the total
number of PEs.

\section{Background}
\label{sec:background}

\subsection{SYCL Programming}
\label{sec:sycl}
SYCL provides a modern generic programming abstraction to enable 
applications on heterogeneous platforms. 
In SYCL, application functions are offloaded to the GPU as parallel
compute kernels. A parallel kernel which operates on thousands or
millions of work-items will be automatically parallelized
and vectorized by the SYCL compiler before being deployed to the GPU.  The
collection of work items are broken into work-groups, that run in
parallel, and sub-groups, that are essentially vector lanes within a
particular compute thread. A work group, depending on hardware, may
consist of 1024 or so parallel work-items, and the overall kernel will
be decomposed into multiple work groups that may run in parallel or
sequentially depending on the GPU compute resources available.

\subsection{oneAPI Level Zero}
\label{sec:level_zero}
oneAPI Level Zero (L0) provides lower layer interfaces for applications 
and libraries to enable programming on accelerators. The API provided by L0
is intended for explicit controls in accelerator programming that is often desired
by higher level runtime APIs and libraries. It provides the back-end support for 
many oneAPI libraries and middleware, such as, oneAPI Threading Building Blocks (oneTBB), 
Intel\regtm{} oneAPI Collective Communications Library (oneCCL), and so on. 

\subsection{OpenSHMEM Memory and Execution Model}
\label{sec:shmem_memory_model}
OpenSHMEM~\cite{openshmem} is a community specification that defines a Partitioned 
Global Address Space (PGAS) parallel programming model. An OpenSHMEM application is 
comprised of multiple Processing Elements (PEs) that execute the same program. Each 
PE is represented by a unique integer value for identification within the range of 
$0\dots npes-1$, where $npes$ denotes the total number of PEs. PEs utilize a 
symmetric data segment and a symmetric heap that can be used for data transfers to 
and from remote PEs. Each PE may contain different values locally in the memory 
objects allocated on symmetric memory locations, however, the layout of the 
symmetric segments is identical at all PEs. This simplifies usage and provides 
opportunities for optimization in access performance. OpenSHMEM defines a library 
API that enables asynchronous, one-sided access to symmetric data at all PEs by 
put and get data transfers, atomic operations, and collective communication 
primitives.

The current OpenSHMEM memory model is specified with respect to the host memory 
being used by a $C$ application. This limits the usage of GPU or other accelerator 
memories as symmetric segments without changes or amendments to the specification. 
The execution model also needs to be adapted to allow all PEs executing at the 
same time with a functional mapping of GPU memories being used as symmetric heap 
by each PE. In SYCL programming model, Unified Shared Memory (USM) allows memory 
from host, device, and shared - that may necessitate different communication and 
completion semantics based on the memory type being used by the runtime. 

%\subsection{Existing OpenSHMEM and GPU Studies}
%\label{sec:related} 

\section{Design and Implementation of Intel SHMEM}
\label{sec:design}

\begin{figure}[b!]
  \centering
  \includegraphics[width=0.65\linewidth]{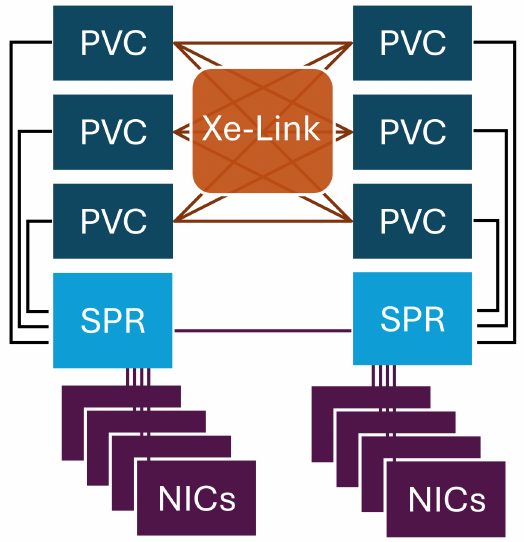}
  \caption{The Aurora compute node architecture.}
  \label{fig:aurora_node}
\end{figure}

\subsection{Architectural Overview}
Intel SHMEM is designed to deliver low latency and high
throughput communications for the compute node architecture of the
Aurora supercomputer system and other tightly coupled GPU systems like it.
A diagram of the Aurora compute node is shown in Figure~\ref{fig:aurora_node},
where 6 Intel\regtm{} Data Center GPU Max Series devices (code named Ponte
Vecchio, or PVC) are fully connected by an Intel\regtm{} Xe-Link fabric.
Figure~\ref{fig:aurora_node} shows a 6-way GPU topology, but Xe-Link can also
be configured with 2-way, 4-way, 6-way, and 8-way topologies, where each GPU
links directly to every other GPU.
The Aurora compute node also features 2 sockets of Intel\regtm{} Xeon\regtm{}
CPU Max (formerly Sapphire Rapids, or SPR) CPUs with optional High Bandwidth
Memory (HBM) attached, PCIe Gen5 CPU-GPU interfaces, and 8 independent
Slingshot 11 Network Interface Card (NIC) devices.

A critical challenge with the distributed programming of such a tightly coupled
GPU system is in simultaneously exploiting the high-speed unified
fabric among the GPUs, while also supporting off-node data communication.
Additionally, SHMEM applications transfer messages with a wide variety of sizes,
so the library must make quick and efficient decisions regarding which path among 
the GPU-GPU fabric, shared host memory, or the network is best for extracting
the most performance possible given the message and the network sizes.
Also, if the data to be transferred is resident in GPU memory, it is
advantageous to use a zero copy design 
rather than to copy the data to host memory and then synchronize the GPU 
and host memory upon data arrival.
GPU-initiated communication is especially important with the advent of GPU
remote direct memory access (RDMA) capabilities, that enable registration of
GPU device memory for zero-copy transfers by the NIC, but present a software
engineering challenge when the feature is unavailable.

\begin{figure}[th!]
  \centering
  \includegraphics[width=0.85\linewidth]{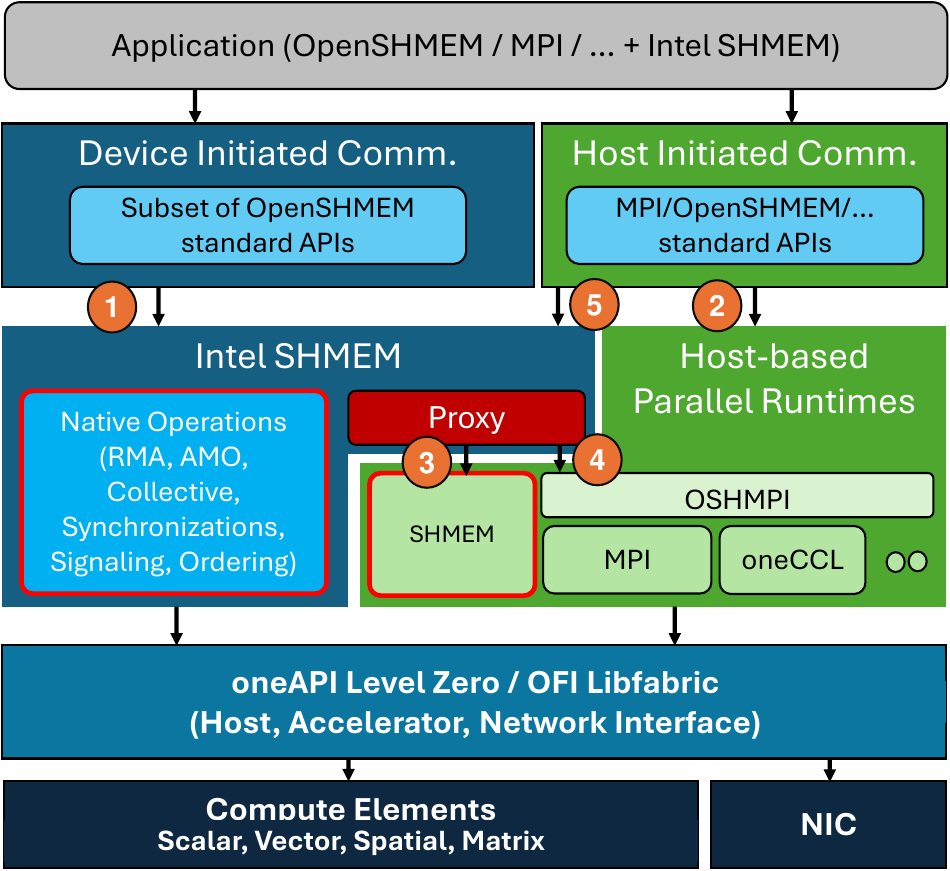}
  \caption{Overview of Intel SHMEM software architecture}
  \label{fig:ishmem_sw_arch}
\end{figure}

With these considerations in mind we present the software architecture for
Intel SHMEM in Figure~\ref{fig:ishmem_sw_arch}.
The top of the figure indicates the application layer, which consists of
Intel SHMEM calls within an OpenSHMEM program (or an MPI program if
interoperability with the host proxy backend is supported, as defined in
OpenSHMEM v1.5 Annex D~\cite{openshmem}).
Intel SHMEM APIs fall into the two broad categories: \textcolor{orange}{(1)}
device-initiated or \textcolor{orange}{(2)} host-initiated (as demarcated by
the light blue or green boxes in Figure~\ref{fig:ishmem_sw_arch},
respectively).
For internode (scale-out) communication, a host-side proxy thread leverages a
standard OpenSHMEM library to hand-off certain GPU-initiated operations,
shown in circle \textcolor{orange}{(3)} within
Figure~\ref{fig:ishmem_sw_arch}.
This proxy backend need not be pure OpenSHMEM; for compatibility with MPI,
the \textcolor{orange}{(4)} OSHMPI solution~\cite{oshmpi} would suffice while
offering competitive performance~\cite{oshmpi-perf}.
As expected, all the \textcolor{orange}{(5)} host-initiated OpenSHMEM
routines are available in Intel SHMEM, with the only caveat being the
routines are prefixed with \texttt{ishmem} as opposed to \texttt{shmem}.
(Similarly, NVSHMEM prefixes with \texttt{nvshmem} and \texttt{ROC\_SHMEM}
with \texttt{roc\_shmem}.)
While this naming scheme may not be strictly necessary, it at least has the
advantage of distinguishing between the various OpenSHMEM-based runtimes that
support GPU abstractions (see Section~\ref{subsec:device-apis}).

\subsection{Xe-Link Capabilities}

Within a local group of GPUs, Xe-Links permit individual GPU threads to issue
loads, stores, and atomic operations to memory located on other GPUs.
Individual loads and stores can provide very low latency.
Many threads executing loads and stores simultaneously can greatly increase bandwidth, but
at the expense of using compute resources (threads) for communications.
To overlap computation and communications, the available
hardware copy engines can be utilized, that can run Xe-Links at full speed while the GPU
compute cores are busy with computation, but at the cost of incurring a
startup latency.

Intel SHMEM, in different operating regimes, uses all these techniques to
optimize performance, doing load-store directly, load sharing among GPU
threads, and using a cutover strategy to use the hardware copy engines for
large transfers and non-blocking operations.
The performance effects of these techniques are presented below in
Section~\ref{sec:performance}.

\subsection{Inter-node and Intra-node Communication}
Much of Intel SHMEM's implementation work has been to support high-performance
GPU-initiated operations for both inter-node (scale-out) and intra-node
(scale-up) communication paths, represented by the dark blue boxes at the
bottom of Figure~\ref{fig:ishmem_sw_arch}.
For inter-node cases, a host-side proxy thread leverages a standard OpenSHMEM
library to hand-off GPU-initiated operations.
Intel SHMEM currently depends on the Sandia OpenSHMEM (SOS)~\cite{sos} for this host
proxy thread backend for two core reasons. First, SOS provides robust and
portable support for all OpenSHMEM v1.5 host-initiated operations by
leveraging the Open Fabrics Interface (OFI) as implemented by
\texttt{libfabric}~\cite{sos-ofi,sos-ctx}.
Additionally, SOS has experimental support for a separate symmetric heap
residing in GPU device memory (see Section~\ref{subsec:device-heap}).
To support RDMA operations on GPU memory, SOS relies on the heterogeneous memory
(\texttt{FI\_HMEM}) capabilities of suitable \texttt{libfabric} providers to
support device memory registration across various high performance
networking interfaces~\cite{hmem}.
For flexibility of usage, Intel SHMEM also provides a choice to use either GPU
memory (default) or host-resident
Unified Shared Memory (USM) as the OpenSHMEM Symmetric Heap.

For intra-node use-cases, Intel SHMEM can directly leverage the Level Zero
inter-process communication (IPC) interfaces without invoking a host proxy
operation~\cite{level-zero}.
The core routine for intra-node transfers is
\texttt{zeCommandListAppendMemoryCopy}, and Intel SHMEM supports both standard
Level Zero command lists and immediate command lists for low latency copy
operations.
Every Intel SHMEM GPU RMA operation first loads from a stashed array to
determine whether the target PE is local.
If this value is non-zero, the target PE is local and the loaded value contains
an appropriate index into an array of \textit{offsets} between the local
symmetric heap base and the destination symmetric heap base.

\subsection{Synchronization with Host}
When a GPU thread encounters an Intel SHMEM operation which requires host
assistance, it composes a request message and transmits it to the host CPU.
If a reply is required, the thread then waits for a completion message.
This reverse offload queue uses a novel lock-free ring buffer design.
The salient features are:

\begin{itemize}
\item Messages are fixed size (64 bytes)
\item Allocation of transmit slots requires a single atomic fetch and
    increment, providing fast arbitration among thousands of GPU threads
\item Message transmission can use a single bus operation
\item Reverse channel flow control is not in the critical path and has less
    than 1\% overhead
\item Completions are independently allocated to permit out of order replies
\item The GPU end does not require a progress thread
\item GPU and CPU communications use only store instructions, that can be
     both fire-and-forget and pipelined
\end{itemize}

The message ring is fast, with about 5\,us round trip time from GPU to
host to GPU, which is close to the required PCIe bus and arbitration times.
Multiple GPU threads can achieve more than 20 million requests per second, even
with only a single thread processing requests at the CPU end.

\subsection{Device-resident Symmetric Heap}
\label{subsec:device-heap}
By design, Intel SHMEM requires a one-to-one mapping of PEs to SYCL devices. 
For Intel Data Center GPU Max Series, each GPU consists of two tiles, both 
of which are considered as a single SYCL device. This 1:1 mapping of PE to 
a GPU tile ensures a separate GPU memory space as symmetric heap for 
the corresponding PE. 
In order for the proxy thread to support host-sided OpenSHMEM operations on
a symmetric heap residing in GPU memory, the buffer must be registered with
the \texttt{FI\_MR\_HMEM} mode bit set~\cite{hmem}.
For Intel SHMEM to indicate which region to register, SOS provides extension
APIs for registering a device memory region as an \textit{external} symmetric
heap, which is supported alongside the standard OpenSHMEM symmetric heap
residing in host memory~\cite{sos}.
These interfaces are:
\begin{itemize}
    \footnotesize{\item \texttt{shmemx\_heap\_create(base\_ptr, size, ...)}}
    \footnotesize{\item \texttt{shmemx\_heap\_preinit()}}
    \footnotesize{\item \texttt{shmemx\_heap\_preinit\_thread(requested, \&provided)}}
    \footnotesize{\item \texttt{shmemx\_heap\_postinit()}}
\end{itemize}
These APIs should be considered experimental, as they are well-suited to SOS's
dual-phase initialization strategy.
In the preinit phase, all PEs allocate their host symmetric heap regions (and
determine the starting address of the static symmetric data segment) and
initialize a process management interface (PMI)~\cite{pmi} to enable a
key-value store for publishing and retrieving all relevant addresses and
information.
After this point, the application can optionally register the external device
region using \texttt{shmemx\_heap\_create}, passing the desired base address,
the region size, and constants indicating the memory kind
(\texttt{SHMEMX\_EXTERNAL\_HEAP\_ZE} and \texttt{SHMEMX\_EXTERNAL\_HEAP\_CUDA}
are currently supported) and device index. The postinit phase then takes 
care of registration of all the symmetric memory regions with the NIC and 
also completes other required initialization steps.

The PGAS community is actively considering alternative APIs for supporting
external memory regions.
OpenSHMEM researchers and specification committee members had studied
memory kinds and spaces\cite{spaces,shmem-mem-kinds}.
UPC++ supports a memory kinds interface leveraging specific GPU allocator
objects~\cite{upcxx}, which has proven to be a suitable replacement for MPI in
a Kokkos application~\cite{upcxx-kokkos}.
Intel SHMEM plans to eventually embrace a standardized OpenSHMEM interface, and
the extensions described above provide a proof-of-concept for such a capability.

\subsection{Proposed Device APIs}
\label{subsec:device-apis}
Intel SHMEM strives to support all the OpenSHMEM APIs, which at the time of this
writing is OpenSHMEM version 1.5.
Some interfaces must be called from the host only, such as the
initialization and finalization APIs and all memory management APIs, because only
the host is well-suited to setup the data structures that handle CPU/GPU proxy
interactions and to perform dynamic allocation of device memory.
On the other hand, most OpenSHMEM communication interfaces including RMA, AMO,
signaling, collective, memory ordering, and point-to-point synchronization
routines are callable from \textit{both} the host and device with
identical semantics.

However, much like NVSHMEM and ROC\_SHMEM, the nature
of GPU programming necessitates \textit{new} device-only APIs that account for the
massively multi-threaded architecture~\cite{b2,rocshmem}.
For example, GPUs schedule \textit{groups} of threads to better exploit the
underlying SIMD (same instruction multiple data) or SIMT (same instruction
multiple thread) architecture.
If, for example, all threads within a group initiate RDMA operations
simultaneously, there can be substantial contention on NIC resources leading to
drastic performance degradation.
In this case it is more performant for the library to restrict RDMA operations
to only occur on a designated leader thread of the thread group.
On the other hand, device-specific APIs could enable threads within a group to
collectively and collaboratively participate in communication operations, for
instance by having each thread copy a given chunk of the source data to the
destination across a unified memory space.

For these reasons, Intel SHMEM includes device-only extensions called the
work\_group interfaces, that are prefixed with \texttt{ishmemx\_} as opposed to
\texttt{ishmem\_}.
Examples of the \code{work\_group} routines for RMA include:
\begin{itemize}
    \item \small{ \texttt{ishmemx\_put\_work\_group} }
    \item \small{ \texttt{ishmemx\_get\_work\_group} }
    \item \small{ \texttt{ishmemx\_put\_nbi\_work\_group} }
    \item \small{ \texttt{ishmemx\_get\_nbi\_work\_group} }.
\end{itemize}
The collective operations also have \code{work\_group} variants, e.g.:
\begin{itemize}
    \item \small{ \texttt{ishmemx\_broadcast\_work\_group} }
    \item \small{ \texttt{ishmemx\_reduce\_work\_group} }
    \item \small{ \texttt{ishmemx\_barrier\_all\_work\_group} }
    \item \small{ \texttt{ishmemx\_sync\_all\_work\_group} }.
\end{itemize}
The AMOs do not have work\_group variants because they are scalar operations
that would not benefit from group optimizations.
All device APIs are listed in ``The \texttt{work\_group} APIs Overview'' section
and documented throughout the ``Intel\regtm{} SHMEM API'' section of the
online specification~\cite{ishmem-spec}.

%\textcolor{red}{TODO: highlight the differences w/ NVSHMEM/ROC\_SHMEM: pt2pt sync and barrier/sync for example, and why}

\subsection{Implementation Choices}

\subsubsection{Remote Memory Access}

The implementation of Intel SHMEM requires careful attention to the presence of independent 
memory systems for the host CPU and for the attached GPU. The SYCL memory model and 
Level Zero GPU runtime system permit mapping of host memory into the GPU address space, 
and mapping of GPU memory into the host address space. 
However, individual loads and stores over PCIe depend on the PCIe bus bandwidth whereas, 
the host-initiated copy engines suffer from the startup cost needed for each transfer.
%but while the PCI express bus 
%provides about 50 GB/s bandwidth, individual loads and stores may take several 
%microseconds.  Bulk data copies can use either load/store or the hardware copy engines. 
%The copy engines achieve high bandwidth, but have 7 or 8 microsecond startup costs and can only be started from the host.
Because of this, Intel SHMEM maintains separate data structures in host memory for 
initialization and host-initiated operations, and in GPU memory for 
GPU-initiated operations.  The GPU data structures are accessible via a SYCL global variable.

Within a node (or potentially within a supernode), GPUs can directly talk to each other 
via Intel Xe-Link.  These links permit load-store, copy engine, and atomic operations.  
During initialization, Intel SHMEM sets up memory mapping from every GPU to the 
symmetric heaps of every other GPU on the local node. Each OpenSHMEM operation then 
is translated to a direct data transfer from one GPU memory to the other. 

As an example, for \code{ishmem\_long\_p(long *dest, long value, int] pe)}, which is a
blocking store of a scalar value to a symmetric address on another processing
element, the implementation works as follows:
\begin{itemize}
\item Base address of GPU resident data structures are loaded
\item A table lookup determines whether the target PE's heap is accessible via store
\item Local address of the symmetric heap on the target PE in loaded for PEs within the node
\item Pointer to \code{dest} is calculated given the target PE: (\code{dest - local\_heap\_base + remote\_heap\_base})
\item Value stored to the \code{dest} buffer on the target PE
\end{itemize}
In case of inter-node when the target PE is not directly accessible, a
reverse offload message is created capturing the parameters of the call and
then passed to the host via the ring buffer.
A host thread receives the message and executes it as though it had been a
host-initiated operation.

For remote memory access (\textit{put} and \textit{get}) that are blocked or
strided, the intra-node version of the operations uses special \code{memcpy}
functions that are based on SYCL vector operations for efficiency.

For the device extension operations mentioned in Section~\ref{subsec:device-apis}, 
the intra-node versions use a multi-threaded vectorized
\code{memcpy} while the inter-node versions perform a SYCL group barrier to
assure the input buffers are valid, and the group leader thread is selected to
make the reverse offload call to the host.

%For both single-threaded and work-group versions of block RMA
%operations, we have implemented cutover logic to switch from the use
%of organic load-store for smaller operations, to, for larger
%oprations, making an upcall to the host in order to start the copy
%engines.  Cutover tuning is dependent on block size and on the number
%of GPU threads active.  Figure~\ref{fig:ishmem_put_bw_3} shows cutover behavior when the
%switching point is size dependent but not thread dependent.  Cases
%with larger numbers of threads should have a higher crossover.  A
%refinement of this design would copy locally enough of the buffer so
%that the GPU load store operations cover the copy engine startup
%latency.

%% Fix this by generating ALWAYS and NEVER cutover graphs

%\begin{figure}[th!]
%  \centering
%  \label{fig:ishmem_put_bw_3}
%  \includegraphics[width=0.85\linewidth]{graphs/p_put_bw_device_3.pdf}
%  \caption{Multi-thread put over Xe-Link.}
%\end{figure}

\vspace{2mm}
\subsubsection{Collective Operations}

Collective operations are often implemented by dynamic switching among
different algorithms depending on interconnect topology, number of ranks, and
operation size for efficiency reasons.
\code{MPI\_Allreduce}, and \code{shmem\_reduce} are good examples.
A small data reduction is usually latency limited, but a large reduction is
usually bandwidth limited and different algorithms are appropriate for 
different data size and configuration choices.

Algorithm selection is particularly relevant to Intel SHMEM, which supports
interconnect aware algorithms for intra-node collectives, and relies on
OpenSHMEM for inter-node operations.

\noindent\textbf{Sync and Broadcast}:
In OpenSHMEM, \textit{sync} is a collective synchronization operation (similar
to \textit{barrier} but without a \textit{quiet}).
\code{ishmem\_team\_sync (ISHMEM\_TEAM\_SHARED)} is the idiom for synchronizing
all the processing elements within the same shared memory domain.
In the case of a system like Aurora, there are 12 GPU tiles per node, and all
of the GPU tiles can be directly accessed for load-store from each other.
After experimenting with different atomic operations, we choose to implement
\textit{sync} by having each PE to send an atomic increment to other PEs on a 
pre-allocated device memory region, and then
waiting locally for the local variable to reach the correct total.
The reason this works is that the Xe-Links can handle a large number of
pipelined remote atomics, that are fire-and-forget, and then the local wait
(implemented by an atomic compare exchange) can use the local GPU caches
effectively.

We use the same ``push'' idea for smaller \textit{broadcast} and
\textit{collect} (which is like MPI \textit{allgather}) operations, because
generally stores are faster than loads, and by having the inner loop of a
broadcast across different destinations, with the outer loop across addresses
we can effectively load share across all the Xe-Links available.

%\begin{figure}[th!]
%  \centering
%  \label{fig:fcollect_bw_12}
%  \includegraphics[width=0.85\linewidth]{graphs/p_fcollect_bw_device_grp1_12.pdf}
%  \caption{12-rank fcollect (AllReduce).}
%\end{figure}

\noindent\textbf{Reduction}:
Some of the more interesting OpenSHMEM collectives are the various
\textit{reduction} operations.
OpenSHMEM includes \textit{min}, \textit{max}, \textit{sum}, \textit{product}, 
\textit{and}, \textit{or}, and \textit{xor} operations for
fixed-point datatypes from 8 to 64 bits and \textit{min}, \textit{max}, 
\textit{sum}, and \textit{product} for
floating-point datatypes.
Since hardware supported atomic operations do not cover all of these datatypes,
we could not adopt the ``push'' strategy as discussed above.
Instead, we exploit the enormous parallelism available on the GPU to split the
reduction by address across threads, and have each thread use vector load
operations, one local and one remote, to assemble the data followed by vector binary
operations to do the reduction, and then vector based stores to update the destination
buffer.
Each PE duplicates the computation, which avoids extra synchronization among
PEs.

For very large reductions, it is a better approach to split the work by address
across PEs and then exchange results, that is the approach taken by the usual
dissemination and ring reduction algorithms.

%\subsubsection{Tests and Launching Applications}
%For completeness we provide a comprehensive set of unit tests and performance
%tests covering device-initiated operations.
%To support different platform environments, we provide a launcher script to
%start Intel\regtm{} SHMEM applications on the available SYCL devices in the
%system with suitable device mappings.

\section{Performance Evaluation}
\label{sec:performance}
In this section, we present performance data for Intel SHMEM operations using
micro-benchmarks implemented to measure minimum latency and maximum bandwidth
achieved on systems with Intel Data Center Max GPUs.
To collect these performance data, an internal system called
Borealis~\cite{borealis} is used with a very similar configuration to that of
Aurora~\cite{aurora}.
Each compute node contains two Intel\regtm{} Xeon\regtm{} CPU Max 9470C
(Sapphire Rapids) CPUs at 2.0 GHz, each with 52 cores and 2 threads per core.
Each node has 1.1\,TB of memory, six Intel\regtm{} Data Center Max 1500 (Ponte
Vecchio) GPUs and are connected using a HPE\othertm{} Slingshot fabric 2.0.2.
Nodes are running SUSE Linux\othertm{} Enterprise Server 15 SP4 with Cray
Shasta kernel 5.14.21-150400.24.55. The six GPUs on each node are fully
connected by an Intel\regtm{} Xe-Link network.

\begin{figure*}[t!]
\begin{center}
\mbox{
    \subfigure[Intel SHMEM Put Bandwidth]
    {
        \label{fig:put_bw}
        \includegraphics[width=0.45\textwidth]{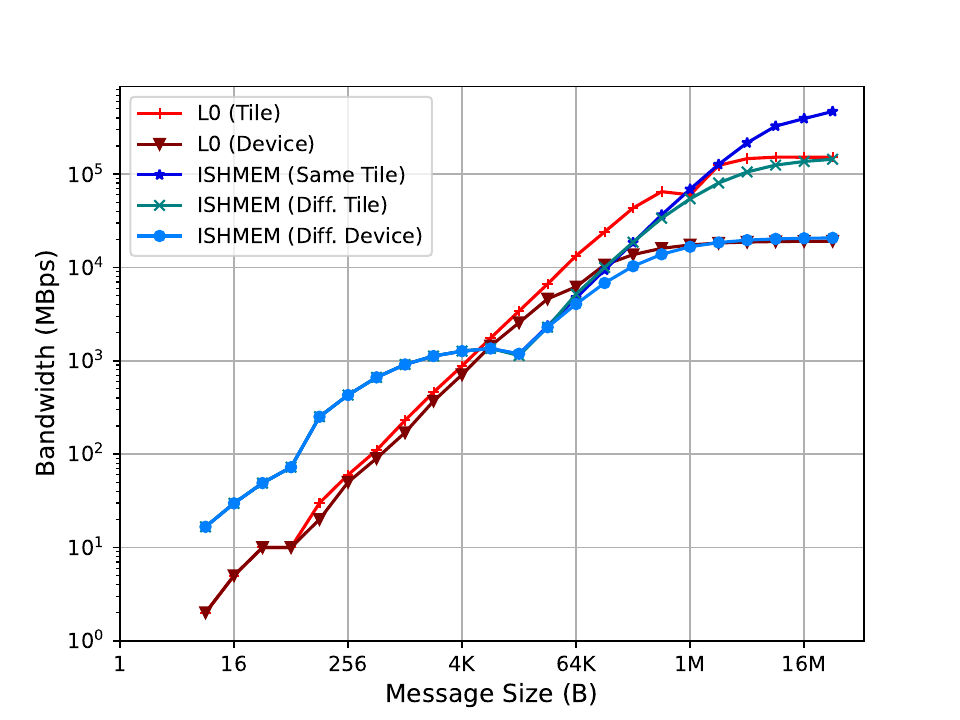}
    }
    \subfigure[Intel SHMEM Get Bandwidth]
    {
        \label{fig:get_bw}
        \includegraphics[width=0.45\textwidth]{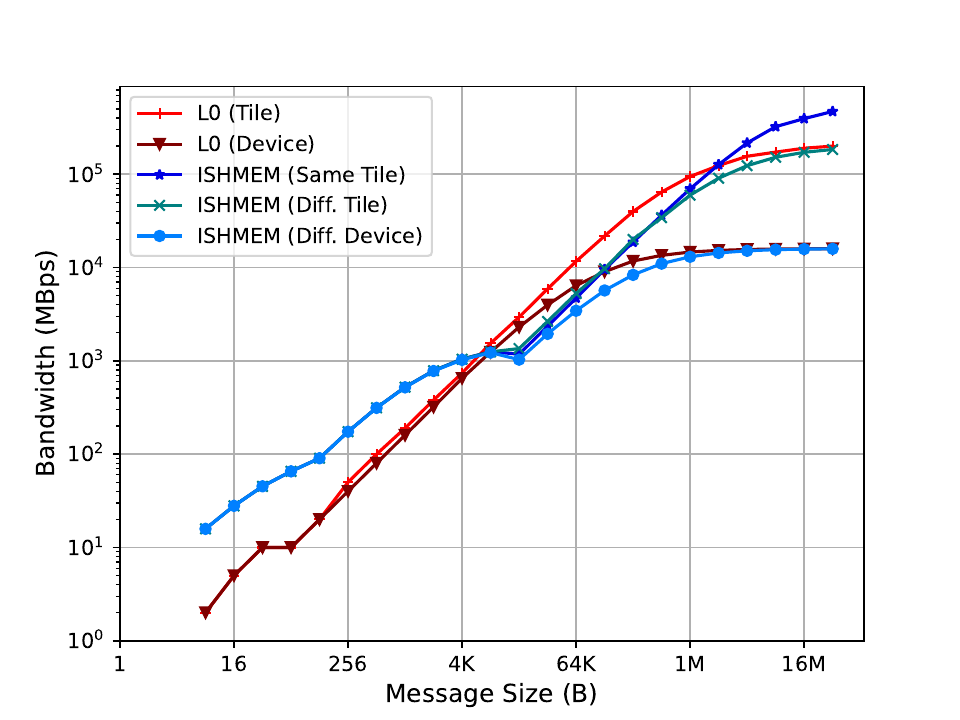}
    }
}
\caption{Intra-node single-threaded Put and Get operation bandwidth within the same device, across tile, and different device}
\label{fig:put_get_bw}
\end{center}
\end{figure*}

As performance measurement methodology, we have used the SYCL profiling
utility~\cite{sycl-profile} by adding the
\code{sycl::property::queue::enable\_profiling()} property to the SYCL queue
that launches the operations.
To reduce performance noise as much as possible, we warm-up the execution by
running a variable number of iterations at the beginning bounded by the total
execution time.
We double the number of iterations until the execution time reaches more
than 2\,ms, at which point we stop the warm-up iterations.
Then, we execute 10 trial iterations and take the best execution time from
these.

In the first set of experiments, we measure the Put and Get operation bandwidth
using the single threaded Intel SHMEM operations, \code{ishmem\_put} and \code{ishmem\_get}.
Figure~\ref{fig:put_get_bw} presents these results.
To compare and contrast the Intel SHMEM performance, we also measure the read
and write operation throughput using a oneAPI Level Zero (L0) benchmark,
\code{ze\_peer}~\cite{ze-peer}.
\code{ze\_peer} is a performance benchmark to measure communication
bandwidth and latency between two L0 devices in a system.

As shown in Figure~\ref{fig:put_bw} and Figure~\ref{fig:get_bw}, performance varies depending on the 
message size of the operation. For small to medium message sizes of up to 4\,KB, Intel SHMEM 
outperforms the L0 benchmark \code{ze\_peer} because the GPU-resident
Intel SHMEM code
directly execute loads or stores to the target PE, which avoids the
startup latency of the GPU copy engines.  Beyond 4\,KB message size, the copy engine 
based transfer performs better. Above  
a tuned cutover value set internally in Intel SHMEM, the host proxy is used to start the copy engines to do the data 
transfer. For the mid-size messages above the cutover, Intel SHMEM performs slightly
worse than L0 due to the extra latency of the reverse offload.  As message size grows beyond 1\,MB, Intel SHMEM 
performs similar to that of L0.  

The graphs also demonstrate performance for three hardware interconnections.
With a single PE execution, the Intel SHMEM performance benchmarks use the
\code{src} and \code{dest} locations on the same GPU tile.
With two PEs, the target PE is on the other tile of the same GPU, and with
three PEs, the target PE is on a different GPU.
Thus, the three different executions of Intel SHMEM benchmarks on
Figure~\ref{fig:put_bw} and Figure~\ref{fig:get_bw} exhibit the performance
characteristics across different hardware components.
%for which, the limits are about 1 TBps,
%153 GBPS, and 20 GBps, respectively.
%Intel SHMEM achieves performance close to the limits of Level Zero.
  
\begin{figure*}[t!]
\begin{center}
\mbox{
    \subfigure[Store from kernel work-items]
    {
        \label{fig:put_cutover_never}
        \includegraphics[width=0.45\textwidth]{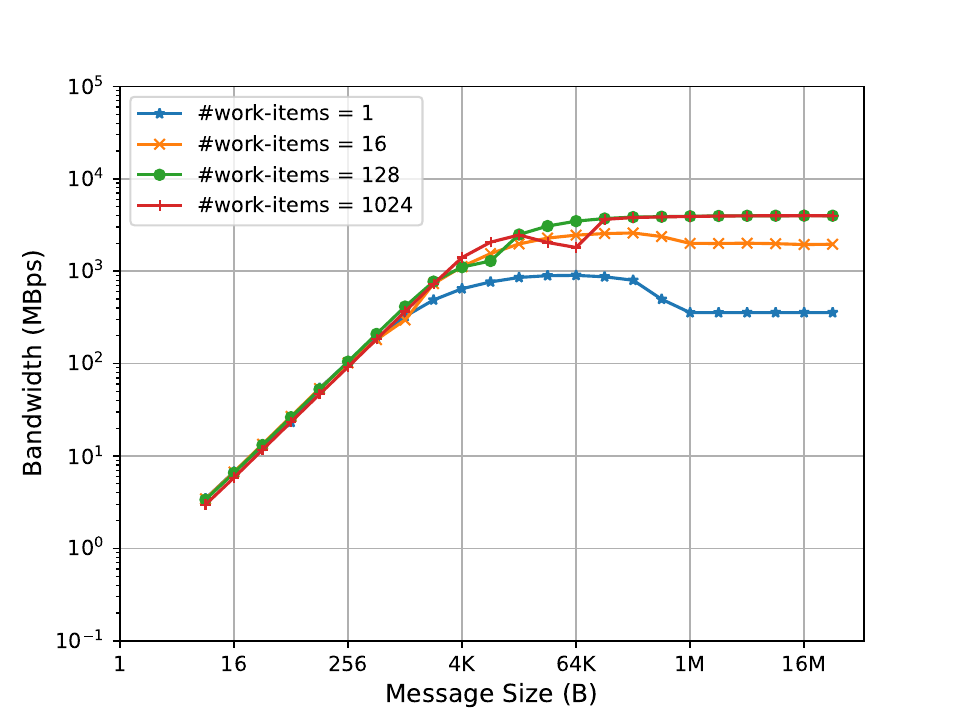}
    }
    \subfigure[Copy engine]
    {
        \label{fig:put_cutover_always}
        \includegraphics[width=0.45\textwidth]{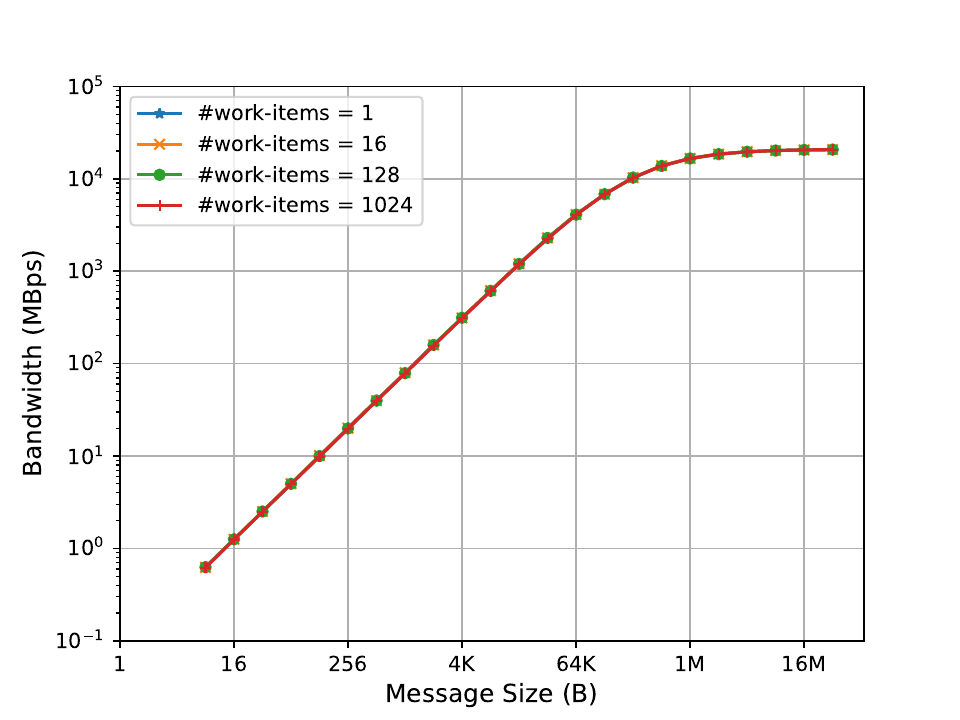}
    }
}
\caption{Performance characteristics of work-group Put operation with varying number of work-items}
\label{fig:put_bw_wg}
\end{center}
\end{figure*}

In the second set of experiments, we measure the bandwidth for the device
extension APIs of RMA, as discussed in Section~\ref{subsec:device-apis}.
We use the same performance methodology as the default Put experiment but vary
the number of work-items within the work-group that we use during kernel
launch.
With these experiments, we show the usefulness of the device extension APIs and
how more number of work-items can parallelize and extract better performance.
Figure~\ref{fig:put_bw_wg} presents these results.
In Figure~\ref{fig:put_cutover_never}, we execute kernel-driven store
operations for all message sizes and in Figure~\ref{fig:put_cutover_always}, we
measure the bandwidth for copy engine driven transfers.
Both of these figures have identical Y-axis bounds to facilitate a direct
comparison. These tests are for the case where source and destination
are on different GPUs connected by the Xe-link fabric.

As shown in Figure~\ref{fig:put_cutover_never}, with increasing
work-group size (threads), for the same data size, performance can be
improved as more resources are utilized.
We run this experiment with 1, 16, 128, and 1024 work-items per work-group and
observe higher bandwidth for 1024 work-items compared to others.
On the other hand, in Figure~\ref{fig:put_cutover_always}, we observe the same
performance for different number of work-items.
This is because for reverse offload, Intel SHMEM chooses only a
single GPU work-item to communicate with the host proxy, thereby minimizing
contention.
These graphs also demonstrate that for small to medium message sizes, store
operations within kernels perform better.
As message size increases, choosing the copy engine is more performant even for
these device extension APIs.
However, unlike the cutover value for standard APIs, the cutover value for the
device extension APIs depend on both the message size and the number of
work-items launched in the work-group as Figure~\ref{fig:put_cutover_never}
shows variability in performance with respect to the work-group size. 

\begin{figure*}[t!]
\begin{center}
\mbox{
    \subfigure[Bandwidth]
    {
        \label{fig:put_bw_wg_2}
        \includegraphics[width=0.45\textwidth]{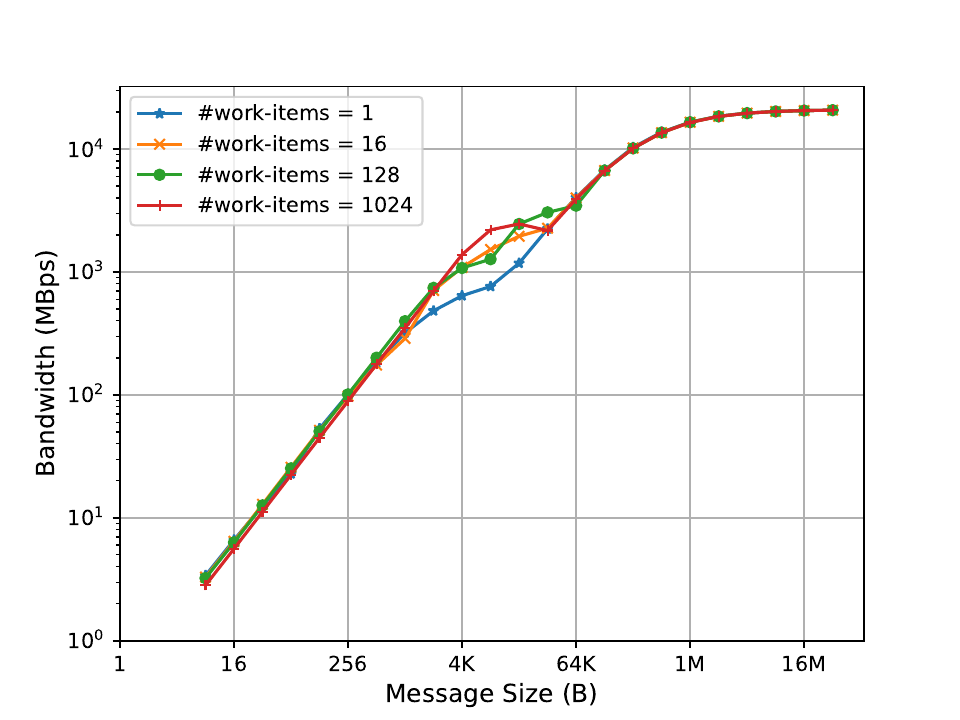}
    }
    \subfigure[Latency]
    {
        \label{fig:put_lat_wg_2}
        \includegraphics[width=0.45\textwidth]{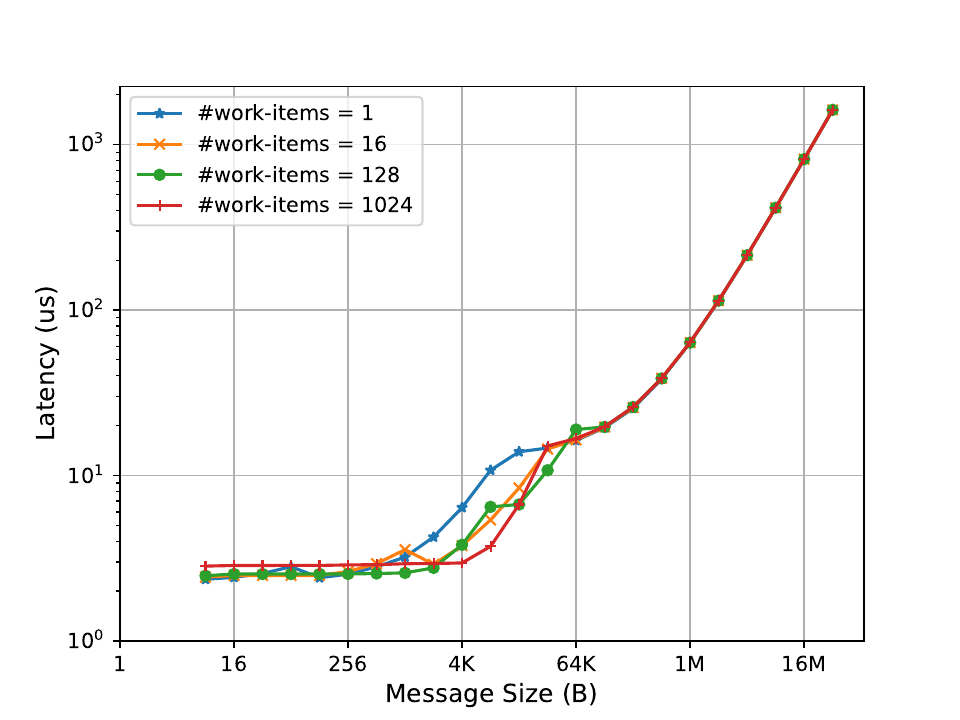}
    }
}
\caption{Performance characteristics of work-group level Put operation with varying number of work-items}
\label{fig:put_bw_lat_wg_2}
\end{center}
\end{figure*}

We, therefore, have implemented cutover logic to switch from the use of organic
load-store for smaller operations, to, for larger operations, making an up-call
to the host in order to start the copy engines.
Cutover tuning is dependent on the data size and on the number of active GPU
work-items.
Figure~\ref{fig:put_bw_lat_wg_2} presents these results in terms of operation
bandwidth (Figure~\ref{fig:put_bw_wg_2}) and latency
(Figure~\ref{fig:put_lat_wg_2}) for \code{ishmemx\_put\_work\_group}.

With cutover value set, \code{ishmemx\_put\_work\_group} obtains better
performance for small to medium message sizes by using direct store using all
work-items in the work-group.
For larger message sizes, after the cutover, it matches the performance as
obtained by the hardware copy engines.

\begin{figure*}[t!]
\begin{center}
\mbox{
    \subfigure[4 PEs]
    {
        \label{fig:fcollect_cutover_4}
        \includegraphics[width=0.32\textwidth]{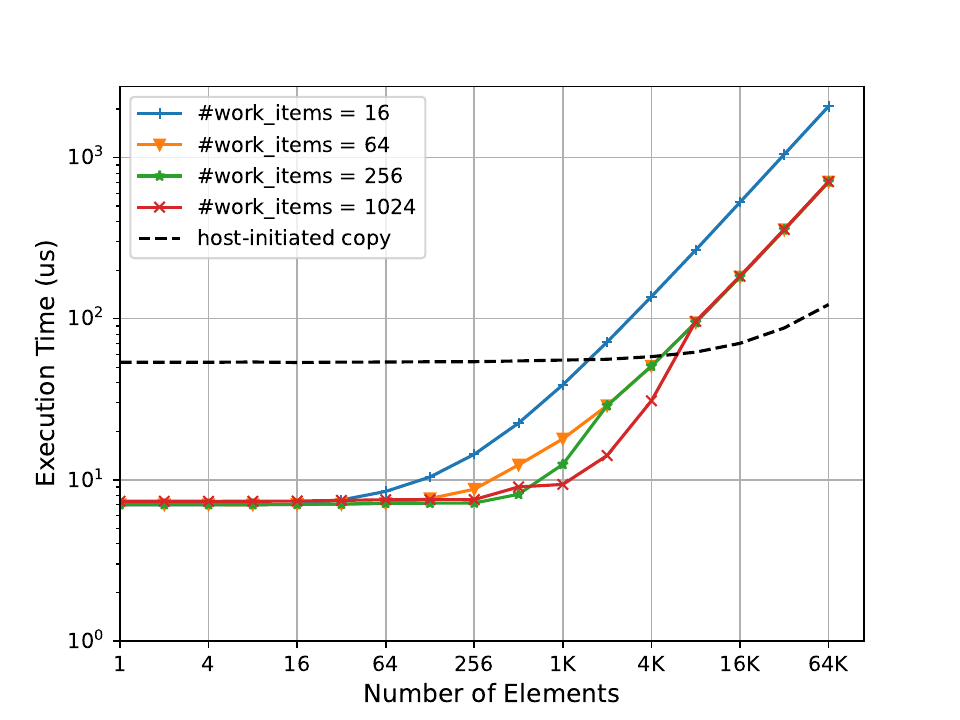}
    }
    \subfigure[8 PEs]
    {
        \label{fig:fcollect_cutover_8}
        \includegraphics[width=0.32\textwidth]{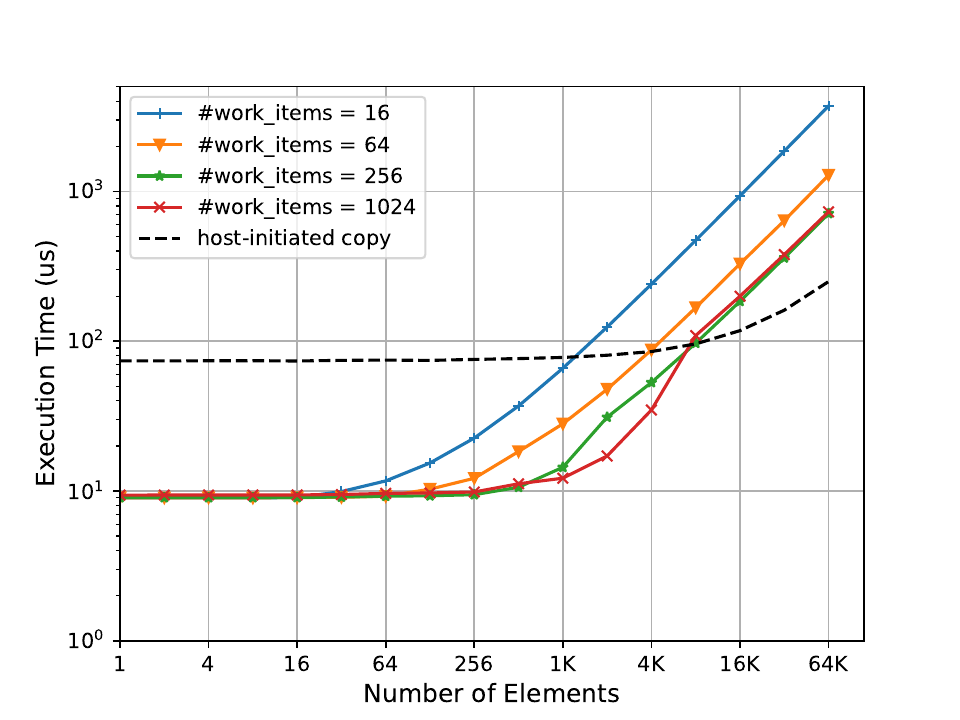}
    }
    \subfigure[12 PEs]
    {
        \label{fig:fcollect_cutover_12}
        \includegraphics[width=0.32\textwidth]{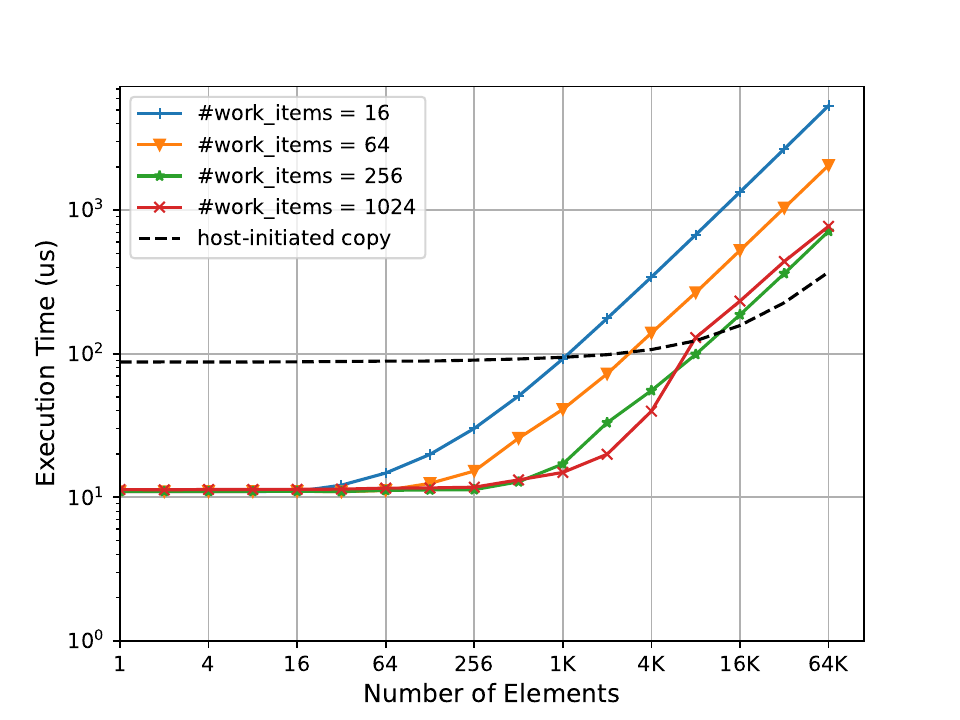}
    }
}
\caption{Performance characteristics of \code{fcollect\_work\_group} with varying number of work-items}
\label{fig:fcollect_cutover}
\end{center}
\end{figure*}

We performed similar experiments for the collective operations and
Figure~\ref{fig:fcollect_cutover} presents the findings for the device
extension of \code{fcollect\_work\_group}.
In this set of experiments, we vary both the number of work-items per
work-group and number of PEs that execute the collective operation.
In these graphs, we present the \textit{host-initiated copy engine} scenario as
the black dashed line to compare with the other device-initiated store
operations.

As shown in Figure~\ref{fig:fcollect_cutover_4},
Figure~\ref{fig:fcollect_cutover_8}, and Figure~\ref{fig:fcollect_cutover_12},
the kernel-initiated direct store operation by the parallel work-items performs
better compared to the host-driven copy engine based transfer for small to
medium number of elements \code{nelems} (around up to 1K).
For larger numbers of \code{nelems}, it is better to choose the reverse offload
for a copy engine transfer.
However, both the number of work-items and the total number of PEs involved in
the collective impact the cutover point.
For example, with 4 PEs and 256 work-items
(Figure~\ref{fig:fcollect_cutover_4}), the cutover point is at 4K number of
elements in the collective.
However, with 12 PEs and 256 work-items (Figure~\ref{fig:fcollect_cutover_12}),
for the same 4K number of elements, it is still better to utilize the parallel
work-items over the host-initiated copy engine transfer.

\begin{figure*}[t!]
\begin{center}
\mbox{
    \subfigure[12 PEs \code{fcollect} performance with tuned cutover]
    {
        \label{fig:fcollect_lat}
        \includegraphics[width=0.45\textwidth]{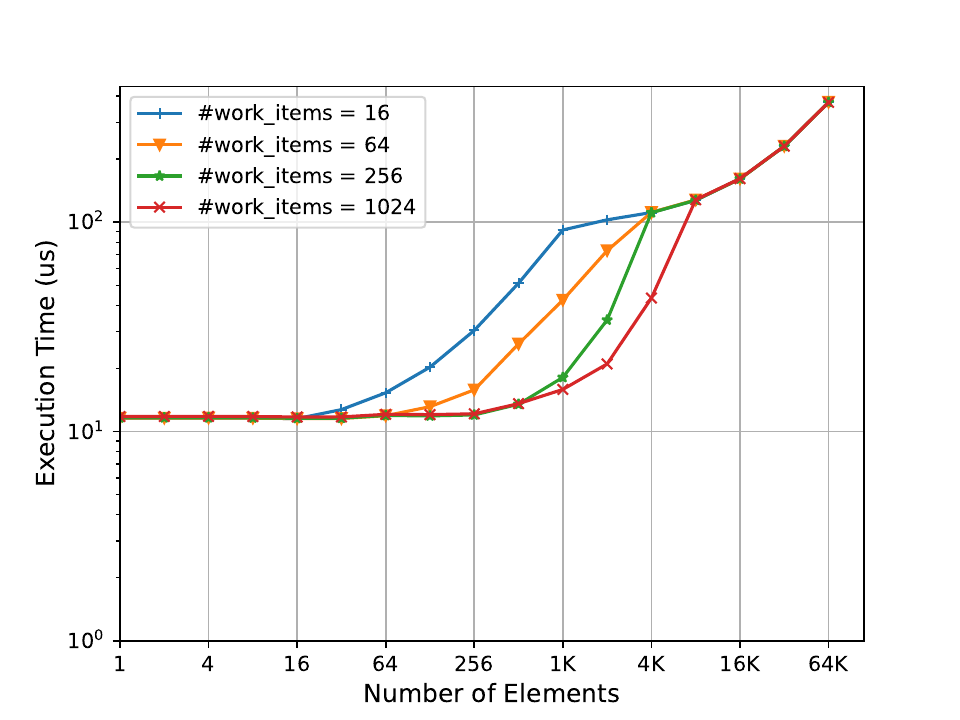}
    }
    \subfigure[\code{broadcast} performance with varying number of PEs]
    {
        \label{fig:bcast_scaleup_lat}
        \includegraphics[width=0.45\textwidth]{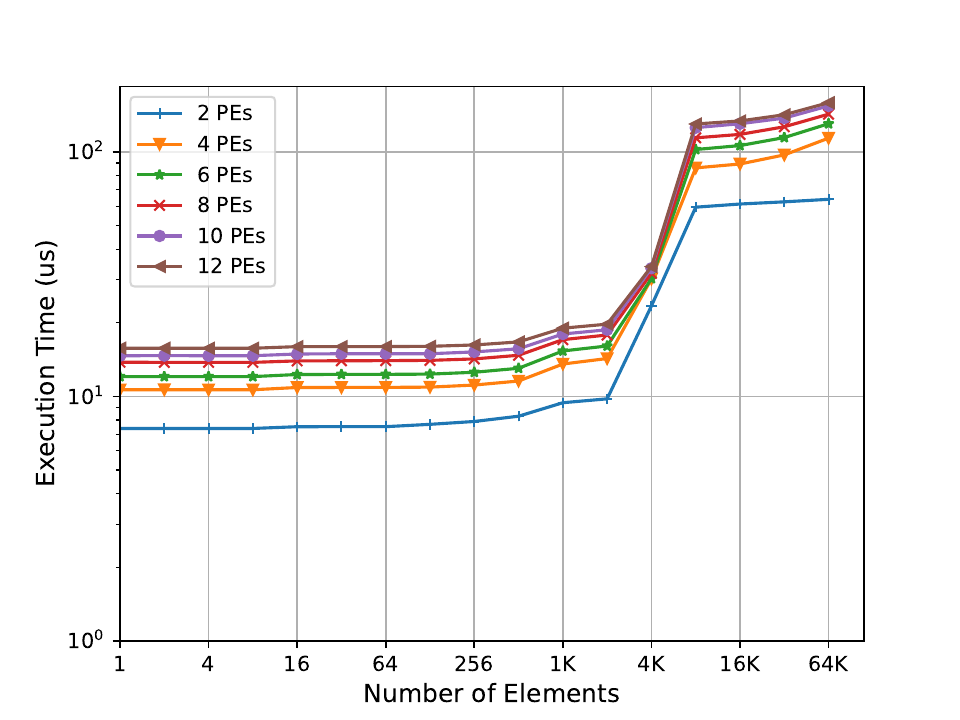}
    }
}
\caption{Performance characteristics of collective operations}
\label{fig:fcollect_bcast}
\end{center}
\end{figure*}

Based on the above findings, we choose the cutover points appropriately for
each collective operation so that the runtime adapts based on the application
execution settings.
We present two collective operations performance in
Figure~\ref{fig:fcollect_bcast}.
Figure~\ref{fig:fcollect_lat} shows performance results for
\code{fcollect\_work\_group} run with 12 PEs and varying number of work-items,
where it appropriately chooses the cutover to provide optimal performance
across different number of elements.
Figure~\ref{fig:bcast_scaleup_lat} presents performance results for
\code{broadcast\_work\_group}, where we vary the number of PEs from 2 to 12,
keeping the number of work-items fixed at 128.
Here, we see uniform strong scaling performance as we increase the number of
PEs.
The performance for 2 PE broadcast stands out as the two PEs in this experiment
are using two tiles within the same GPU, avoiding any across Xe-Link transfer.

\subsection{Comparison with Existing Runtimes}
\label{sec:comparison}
Because of unavailability of systems that contain GPUs from different vendors
and also the lack of back-end support for enabling different GPU drivers, we
could not conduct any experiment comparing performance of Intel SHMEM with that
of NVSHMEM~\cite{b2,b1} and ROC\_SHMEM~\cite{rocshmem}.
As we envision Intel SHMEM providing the non-proprietary solution for
GPU-initiated communication, we would continue working on enabling our library
on other platforms.

\section{Conclusion}

Intel SHMEM is an implementation of the OpenSHMEM standard for the 
C/C++ programming environment with SYCL, initially supporting 
Intel CPUs and GPUs. The current open-source version~\cite{ishmem} 
supports remote memory access, remote atomic operations, signaling, 
synchronization, collective operations, and ordering from OpenSHMEM 1.5 
callable from either
host or device using SYCL kernels, providing coverage to
most SHMEM operations from the entire application.

Extensions to the OpenSHMEM standard are also provided callable from device, in
order to permit heavily multithreaded GPU kernels to collaboratively
implement data transfer operations. Intel SHMEM optimizes by tuning 
these operations appropriately to choose either the kernel-driven load/store 
or the host driven copy depending on the application configurations.

As of next steps, we plan to investigate further on other extensions 
that enable further control for application programmers to define 
dependencies between host and device executions. We also plan to enable 
Intel SHMEM for different application use-cases.

%\section*{References}

\bibliographystyle{IEEETranS}
\bibliography{references}

~\\
\scriptsize
\framebox{
  \parbox{\linewidth}{
  Intel and Xeon are trademarks of Intel Corporation in the U.S. and/or other
  countries.

  \vspace{0.5em}
  Benchmark results were obtained prior to implementation of recent software
  patches and firmware updates intended to address exploits referred to as
  ``Spectre'' and ``Meltdown''. Implementation of these updates may make
  these results inapplicable to your device or system.

  \vspace{0.5em}
  Software and workloads used in performance tests may have been optimized for
  performance only on Intel\regtm{} microprocessors. Performance tests, such as
  SYSmark\othertm{} and MobileMark\othertm{}, are measured using specific
  computer systems, components, software, operations and functions. Any
  change to any of those factors may cause the results to vary. You should
  consult other information and performance tests to assist you in fully
  evaluating your contemplated purchases, including the performance of that
  product when combined with other products.

  \vspace{0.5em}
  For more information go to \url{http://www.intel.com/benchmarks}.
  }
}
~\\
\scriptsize\noindent\othertm{}Other names and brands may be claimed as the
property of others.

\end{document}

\twocolumn[%
{\begin{center}
\Huge
Appendix: Artifact Description/Artifact Evaluation        
\end{center}}
]

%%%%%%%%%%%%%%%%%%%%%%%%%%%%%%%%%%%%%%%%%%%%%%%%%%%%%
%  AD Appendix
%%%%%%%%%%%%%%%%%%%%%%%%%%%%%%%%%%%%%%%%%%%%%%%%%%%%%

\appendixAD

\section{Overview of Contributions and Artifacts}

\subsection{Paper's Main Contributions}

\artexpl{
%Provide a list of all main contributions of the paper.

}

%\artsampl{
\begin{description}
\item[$C_1$] Intra-node single-threaded OpenSHMEM RMA operation bandwidth within different topology
\item[$C_2$] Performance characteristics of OpenSHMEM RMA extensions on GPUs
\item[$C_3$] Performance characteristics of OpenSHMEM collective extensions on GPUs
\end{description}
%}

\subsection{Computational Artifacts}
\label{sec:artifacts}

\artexpl{
%List the computational artifacts related to this paper along with their respective DOIs. Note that all computational artifacts may be archived under a single DOI.
}

%\artsampl{
\begin{description}
\item[$A_1$] https://doi.org/10.5281/zenodo.13749597
\end{description}
%}

\artexpl{
%Provide a table with the relevant computational artifacts, 
%highlight their relation to the contributions (from above) and 
%point to the elements in the paper that are reproducible by each artifact, e.g., 
%which figures or tables were generated with the artifact.
}

%\artsampl{
\begin{center}
\begin{tabular}{rll}
\toprule
Artifact ID  &  Contributions &  Related \\
             &  Supported     &  Paper Elements \\
\midrule
$A_1$   &  $C_1$ & Figure 3 \\
\midrule
$A_1$   &  $C_2$ & Figures 4,5 \\
\midrule
$A_1$   &  $C_3$ & Figures 6,7 \\
\bottomrule
\end{tabular}
\end{center}
%}

%%%%%%%%%%%%%%%%%%%%%%%%%%%%%%%%%%%%%%%%%%%%%%%%%%%%%%%%
\section{Artifact Identification}
%%%%%%%%%%%%%%%%%%%%%%%%%%%%%%%%%%%%%%%%%%%%%%%%%%%%%%%%

\artexpl{
%Provide the following six subsections for each computational artifact $A_i$.
}

\newartifact

\artrel

\artexpl{
%    Briefly explain the relationship between the artifact and contributions.
}

The artifact $A_1$ includes the necessary scripts to generate all three major contributions 
of the paper as listed above. To make it simple to use, separate scripts are provided for 
generating the dataset and graphs. Also, some dataset generation requires the software to 
be re-built with provided patches in $A_1$. For each such dataset, separate script is 
provided.

\artexp

\artexpl{
%Provide a higher level description of what outcome to expect from the corresponding experiments. Provide an explanation of how the results substantiate the main contributions.
}

Each experiment generates a dataset or graph which will show performance characteristics 
of different Intel\regtm{} operations. For $C_1$, the dataset should show better performance 
for \textit{Same Tile} or \textit{Diff. Tile} topology compared to \textit{Diff. Device}. 
For $C_2$, the experiments were run with different number of work items where larger number 
of work-items are expected to perform better compared to the smaller number of work-items.
For $C_3$, along with varying work-items the total number of processes run are also varied 
and these should exhibit better scaled performance with more number of processes.

%\artsampl{
%Algorithm A should be faster than Algorithms C and B in all GPU scenarios.    
%}

\arttime

\artexpl{
%Estimate the time required to reproduce the artifact, providing separate estimates for the individual steps: Artifact Setup, Artifact Execution, and Artifact Analysis.
}

%\artsampl{
The artifact setup would require an estimated time of 5 - 10 minutes approximately, whereas for 
execution of each of the scripts would take 3 - 5 minutes. 
%}

\artin

\artinpart{Hardware}

%\artexpl{
%Specify the hardware requirements and dependencies (e.g., a specific interconnect or GPU type is required).
To run the experiments, the following hardware setup is required.
\begin{itemize}
\item Intel\regtm{} Data Center Max 1500 (Ponte Vecchio) GPU
\item Intel\regtm{} Xeon\regtm{} CPU Max 9470C (Sapphire Rapids) CPU
\item HPE\othertm{} Slingshot interconnect
\item Intel\regtm{} Xe-Link fabric
\end{itemize}
%}

\artinpart{Software}

%\artexpl{
%Introduce all required software packages, including the computational artifact. For each software package, specify the version and provide the URL.
To run the experiments, the following software are required.
\begin{itemize}
\item Linux OS with dmabuf support (e.g., SUSE Linux Enterprise Server 15 SP4)
\item Intel\regtm{} oneAPI DPC++/C++ Compiler 2024.0 or higher
\item oneAPI Level Zero
\item Libfabric (with CXI provider support for HPE Slingshot)
\item Sandia OpenSHMEM
\item Intel\regtm{} SHMEM
\item Python3 with matplotlib
\item System utility tools: \textit{grep}, \textit{awk}, \textit{paste} 
\end{itemize}
%}

\artinpart{Datasets / Inputs}

%\artexpl{
%Describe the datasets required by the artifact. Indicate whether the datasets can be generated, including instructions, or if they are available for download, providing the corresponding URL.
%}

The datasets are generated using the generate\_c$i$ scripts provided in $A_1$.  

\artinpart{Installation and Deployment}

%\artexpl{
%Detail the requirements for compiling, deploying, and executing the experiments, including necessary compilers and their versions.
%}

For Intel\regtm{} SHMEM installation along with all the dependencies, please follow the 
instructions provided in 
\url{https://github.com/oneapi-src/ishmem/blob/main/README.md}. As part of the installation, 
Intel SHMEM requires Level Zero and Sandia OpenSHMEM to be installed. To build and install 
Sandia OpenSHMEM on HPE Slingshot fabric, please follow the instructions provided in 
\url{https://github.com/Sandia-OpenSHMEM/SOS/wiki/Slingshot-(CXI)-Build-Instructions}.
Installation instructions for Libfabric can be found in 
\url{https://github.com/ofiwg/libfabric/blob/main/README.md}.

To install matplotlib on Python, please follow the instructions provided in 
\url{https://matplotlib.org/stable/install/index.html}. 

\artcomp

\artexpl{
%Provide an abstract description of the experiment workflow of the artifact. It is important to identify the main tasks (processes) and how they depend on each other. 

%A workflow may consist of three tasks: $T_1, T_2$, and $T_3$. The task $T_1$ may generate a specific dataset. This dataset is then used as input by a computational task $T_2$, and the output of $T_2$ is processed by another task $T_3$, which produces the final results (e.g., plots, tables, etc.). State the individual tasks $T_i$ and provide their dependencies, e.g., $T_1 \rightarrow T_2 \rightarrow T_3$.

%Provide details on the experimental parameters. How and why were parameters set to a specific value (if relevant for the reproduction of an artifact), e.g., size of dataset, number of data points, input sizes, etc. Additionally, include details on statistical parameters, like the number of repetitions.
}

The artifact follows a workflow of three steps to generate the contributions that are 
mentioned in the paper. Before executing the following workflow, all the required 
software must be installed and ready to use.

The first step on the workflow is to apply one or more patches on the Intel SHMEM 
source code and rebuild Intel SHMEM along with all the tests. This will ensure some 
of the performance enhancements mentioned in the paper to be enabled.

The second step on the workflow is to generate the dataset using the generate\_ci 
scripts. For $C_2$ and $C_3$, the generate scripts are subdivided to ensure 
ease-of-use and simplicity. All of these scripts must be provided with the Intel 
SHMEM build directory for correct execution.

The third step on the workflow is to use the plot files in $A_1$ to generate graphs. 
For ease-of-use, $A_1$ includes a generate\_graphs.sh script which reads all the plot files 
and generates corresponding graphs. To ensure for these scripts to work properly, all 
generated data files should be kept in the same directory as the plot script files.

\artout

After the graphs are generated, those can be analyzed by comparing with the corresponding 
figures as mentioned in Section~\ref{sec:artifacts}.

%\newartifact

%\artexpl{
%Provide the same type of information as done for Computational Artifact $A_1$.
%}

%%%%%%%%%%%%%%%%%%%%%%%%%%%%%%%%%%%%%%%%%%%%%%%%%%%%%
%  AE Appendix
%%%%%%%%%%%%%%%%%%%%%%%%%%%%%%%%%%%%%%%%%%%%%%%%%%%%%
\newpage
\appendixAE

\arteval{1}
\artin

\artexpl{
%Provide instructions for installing and compiling libraries and code. 
%Offer guidelines on deploying the code to resources.
}

The following commands are used to download, build, and install 
Sandia OpenSHMEM (SOS) and Intel SHMEM.
These instructions are specific to the Sunspot cluster at the Argonne
Leadership Computing Facility (ALCF). 
Changes specific to Sunspot@ALCF are highlighted below, so that the commands can be
more easily modified to run on other systems.
At the time of this writing, these are the (default) loaded modules on Sunspot@ALCF,
seen with the \texttt{module list} command:

\scriptsize{
    \begin{verbatim}
Currently Loaded Modules:
  1) spack-pe-gcc/0.7.0-24.086.0    
  2) gmp/6.2.1-pcxzkau              
  3) mpfr/4.2.0-w7v7yjv  
  4) mpc/1.3.1-dfagrna   
  5) gcc/12.2.0                       
  6) mpich/icc-all-pmix-gpu/20231026  
  7) mpich-config/collective-tuning/1024  
  8) intel_compute_runtime/release/821.36 
  9) oneapi/eng-compiler/2024.04.15.002  
  10) libfabric/1.15.2.0                  
  11) cray-pals/1.3.3
  12) cray-libpals/1.3.3
    \end{verbatim}
}
\normalsize

Also on Sunspot@ALCF, please load the default CMake module, at the time of this
writing \texttt{module load cmake/3.27.7}.

Please note that the \code{mpicc} compiler is used for SOS build because of the
usage of an MPI based PMI interface in our execution environment. 
Also, OFI Libfabric is a dependency of SOS, which is already installed on
Sunspot@ALCF.
If building libfabric from source, please note that the provider must have the
\texttt{FI\_HMEM} feature available to support Intel SHMEM.
At the time of this writing, these providers include \texttt{cxi},
\texttt{psm3}, and \texttt{verbs;ofi\_rxm}, which are set via the
\texttt{FI\_PROVIDER} environment variable.
Level Zero (L0) is a dependency of Intel SHMEM, which is also already installed
on Sunspot@ALCF, as noted below.

Before downloading the artifact $A_1$ repository, the following needs to 
be executed to enable proper HTTP proxy configurations.

\scriptsize{
\begin{verbatim}
export HTTP_PROXY=http://proxy.alcf.anl.gov:3128
export HTTPS_PROXY=http://proxy.alcf.anl.gov:3128
export http_proxy=http://proxy.alcf.anl.gov:3128
export https_proxy=http://proxy.alcf.anl.gov:3128
git config --global http.proxy http://proxy.alcf.anl.gov:3128
\end{verbatim}
}
\normalsize

The following steps assume that artifact $A_1$ is cloned into a directory named
\texttt{artifacts}.

\scriptsize{
\begin{verbatim}
git clone --recurse-submodules \
          https://github.com/Sandia-OpenSHMEM/SOS.git sos
cd sos
./autogen.sh
mkdir build
mkdir install

# Set an SOS installation directory:
export SOS_INSTALL=$PWD/install

# If needed, set OFI and L0 installation directories:
# on Sunspot@ALCF with the modules noted above, please use:
export OFI_INSTALL=/opt/cray/libfabric/1.15.2.0
export L0_PREFIX=/opt/aurora/24.086.0/intel-gpu-umd/821.36

cd build
../configure --prefix=${SOS_INSTALL} \
             --with-ofi=${OFI_INSTALL} \
             --enable-pmi-mpi --disable-fortran \
             --enable-ofi-mr=basic \
             --enable-ofi-manual-progress \
             --disable-libtool-wrapper \
             --disable-bounce-buffers \
             --enable-mr-endpoint --enable-ofi-hmem \
             --disable-ofi-inject --disable-nonfetch-amo \
             --enable-manual-progress CC=mpicc CXX=mpicxx
make -j
make install
cd ../..

# Download artifacts including Intel SHMEM v1.1.0:
git clone \
    git@github.com:wrrobin/ws_pawatm114_reprod.git artifacts

cd artifacts
tar -xvf ishmem-v1.1.0.tar.gz
cd ishmem-1.1.0 
mkdir build
cd build
cmake .. -DCMAKE_BUILD_TYPE=Release -DENABLE_OPENSHMEM=ON \
         -DSHMEM_INSTALL_PREFIX=${SOS_INSTALL} \
         -DL0_INSTALL_PREFIX=${L0_PREFIX}
make -j
cd ../..
cp scripts/*.sh ishmem-1.1.0/build
cp scripts/*.py ishmem-1.1.0/build
\end{verbatim}
}
\normalsize

To generate the figure plots, Python3 with Matplotlib is required and can be
installed on Sunspot@ALCF like this:

\scriptsize{
\begin{verbatim}
python3 -m pip install -U matplotlib
\end{verbatim}
}
\normalsize

\artcomp
After the installation of Intel SHMEM, the generate scripts are used to 
generate the dataset as follows. 

The following executions assume a single-node job with the hardware and
software requirements in \textit{Artifact Setup} section.
On Sunspot@ALCF, such a node can be reserved for 30 minutes like this:

\scriptsize{
\begin{verbatim}
qsub -l select=1 -l walltime=30:00 -A <your_project_name> \
     -q workq -I
\end{verbatim}
}
\normalsize

When executing the following scripts on Sunspot@ALCF, the \texttt{FI\_PROVIDER}
should be set to:

\scriptsize{
\begin{verbatim}
export FI_PROVIDER="cxi,tcp;ofi_rxm"
\end{verbatim}
}
\normalsize

\noindent If \texttt{FI\_PROVIDER} is left unset, the scripts will set it to
\texttt{cxi} by default.

To execute a script that generates the $C_1$ contributions, run:

\scriptsize{
\begin{verbatim}
cd ishmem-1.1.0/build
ISHMEM_BUILD_DIR=$PWD ./generate_c1.sh
\end{verbatim}
}
\normalsize

For $C_2$ contributions, there are additional performance enhancing 
features investigated and implemented on top of the current released version 
Intel SHMEM library. These changes are provided as separate patches to be merged 
before generating the datasets. 

\scriptsize{
\begin{verbatim}
cd .. # ishmem-1.1.0
git apply ../ishmem_common.patch
git apply ../ishmem_cutover_never.patch
cd build 
make -j
ISHMEM_BUILD_DIR=$PWD ./generate_c2_1.sh

cd ..
git apply ../ishmem_cutover_always.patch
cd build
make -j
ISHMEM_BUILD_DIR=$PWD ./generate_c2_2.sh

cd ..
git apply ../ishmem_cutover_current.patch
cd build
make -j
ISHMEM_BUILD_DIR=$PWD ./generate_c2_3.sh
\end{verbatim}
}
\normalsize

For $C_3$ also, the patches need to be applied first before generating the dataset.
One thing to note is that, the patches need to be applied in the same order as shown 
here to ensure successful update to the code and generation of the correct results. 
Also, at the time of this writing, the environment variable 
$ZE\_FLAT\_DEVICE\_HIERARCHY$ might need to be unset before executing the following commands.

\scriptsize{
\begin{verbatim}
cd .. # ishmem-1.1.0
git reset origin/main --hard
git apply ../ishmem_common.patch
git apply ../ishmem_cutover_never.patch
cd build
make -j
ISHMEM_BUILD_DIR=$PWD ./generate_c3_1.sh

cd ..
git apply ../ishmem_cutover_always.patch
git apply ../ishmem_cutover_current.patch
cd build
make -j
ISHMEM_BUILD_DIR=$PWD ./generate_c3_2.sh
\end{verbatim}
}
\normalsize

\artexpl{
%Describe the experiment workflow. 
%If encapsulated within a workflow description or equivalent (such as a makefile or script), clearly outline the primary tasks and their interdependencies. Detail the main steps in the workflow. Merely instructing to “Run script.sh” is inadequate.
}

\artout
After all the data is generated through the generate scripts, the generated data files 
are copied to the artifact $A_1$ repository, where the plot files exist. Alternatively, 
the Python scripts (\code{*.py} files) for generating the plots can be copied to the Intel SHMEM build 
directory. The 
\texttt{generate\_graphs.sh} script is then used to generate all the graphs.

\begin{figure*}[t!]
\begin{center}
\mbox{
    \subfigure[Put Bandwidth]
    {
        \label{fig:put_bw}
        \includegraphics[width=0.45\textwidth]{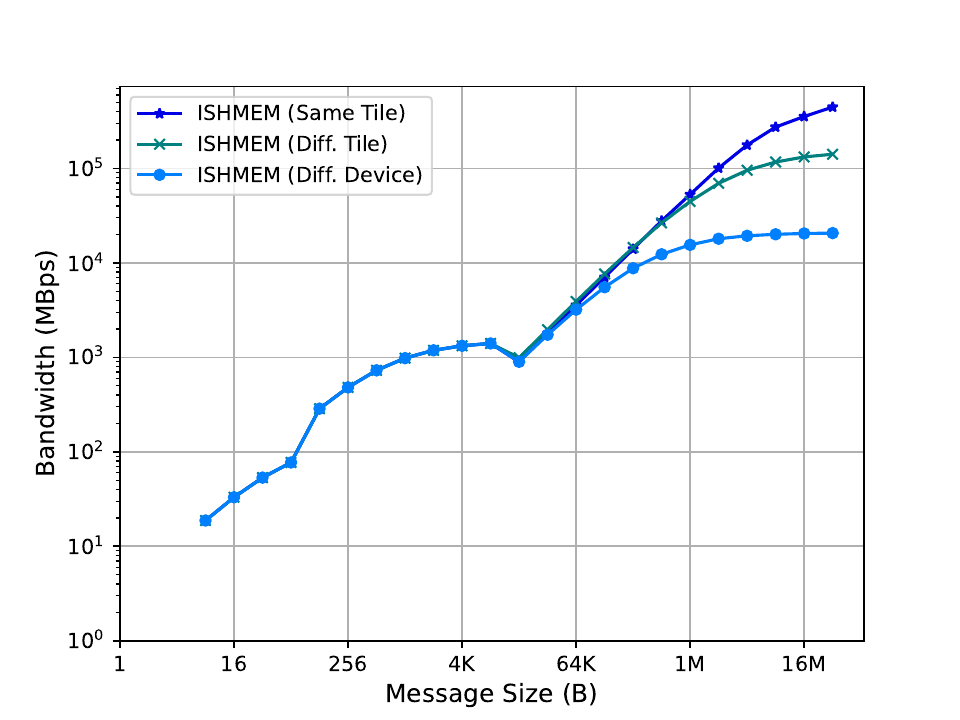}
    }
    \subfigure[Get Bandwidth]
    {
        \label{fig:get_bw}
        \includegraphics[width=0.45\textwidth]{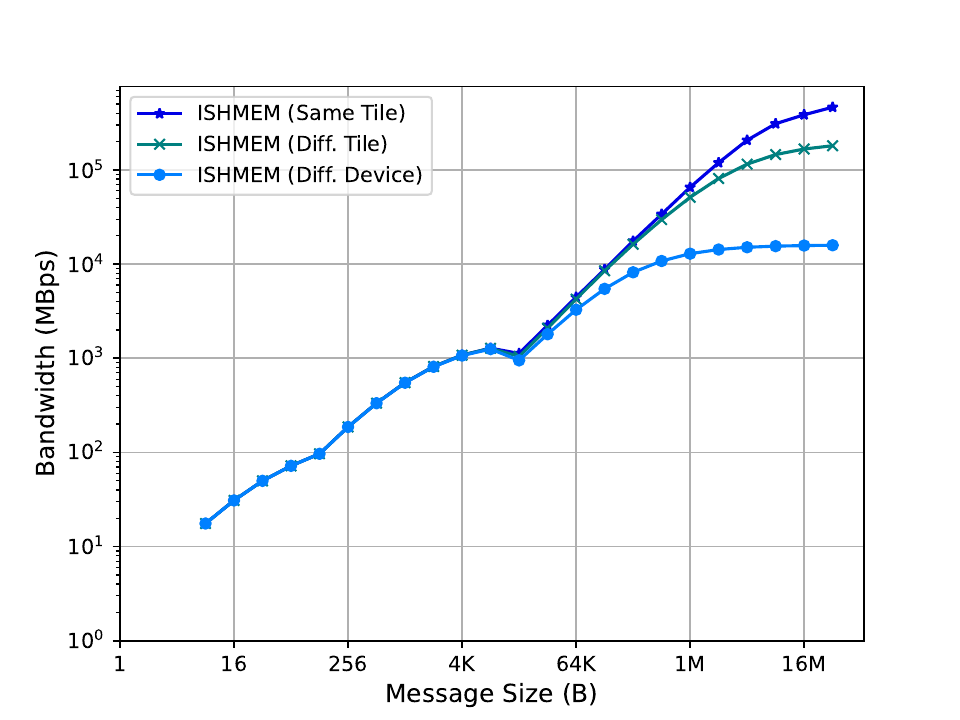}
    }
}
\caption{Intra-node single-threaded Put and Get operation bandwidth within the same device, across tile, and different device}
\label{fig:put_get_bw}
\end{center}
\end{figure*}

Figure~\ref{fig:put_get_bw} presents the reproduced data for Put and Get 
operation bandwidth with Intel SHMEM. Similar to Figure $3$ in the paper, the
data shows similar trend on different topology and also perform best within the 
same GPU tile.

\begin{figure*}[t!]
\begin{center}
\mbox{
    \subfigure[12 PEs \code{fcollect} performance with tuned cutover]
    {
        \label{fig:fcollect_lat}
        \includegraphics[width=0.45\textwidth]{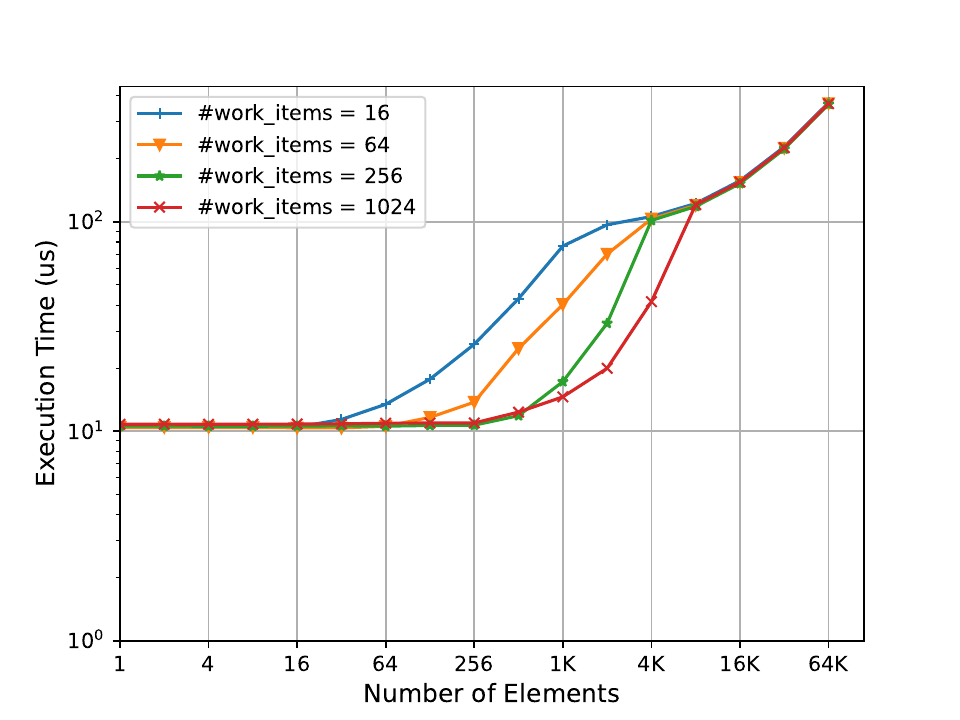}
    }
    \subfigure[\code{broadcast} performance with varying number of PEs]
    {
        \label{fig:bcast_scaleup_lat}
        \includegraphics[width=0.45\textwidth]{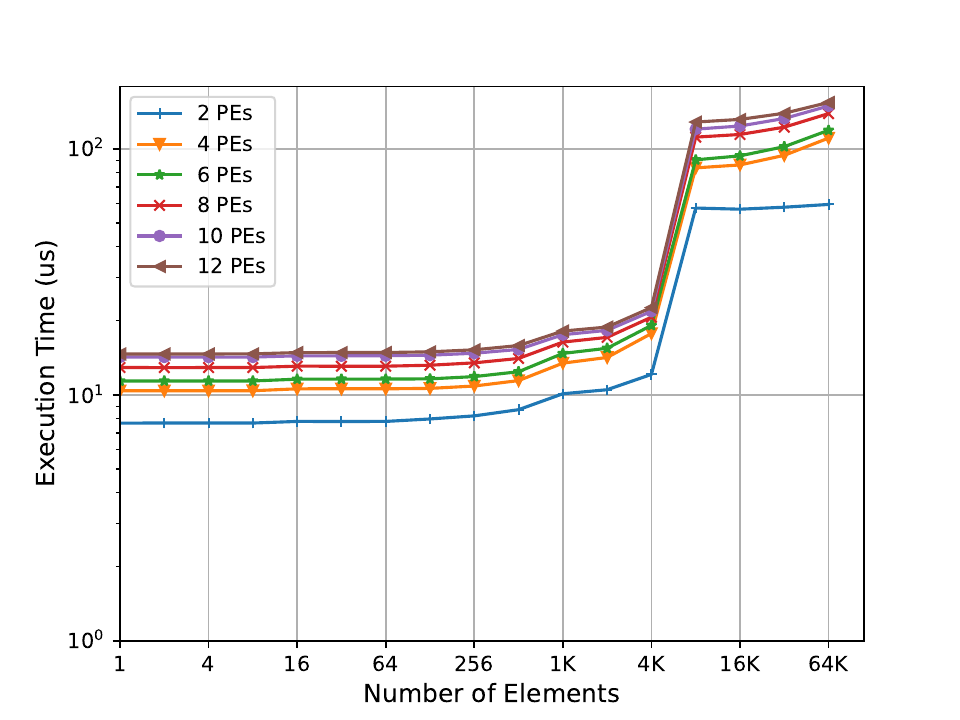}
    }
}
\caption{Performance characteristics of collective operations}
\label{fig:fcollect_bcast}
\end{center}
\end{figure*}

We also reproduce the collective operation performance with fcollect and 
broadcast and present the obtained results in Figure~\ref{fig:fcollect_bcast}.
Similar to Figure $7$ in the paper, the data show good performance 
trend with tuned cutover point selected for switching from kernel initiated EU driven 
copy to host initiated copy engine transfer. 

\artexpl{
%\begin{itemize}
%    \item Provide a description of the expected results and a methodology for evaluating these results. 
%    \item Explain how the expected results from the experiment workflow correlate with the contributions stated in the article. 
%    \item For example, if the article presents results in a figure, the artifact evaluation should also produce a similar figure, depicting the same generalizable outcome. Authors must focus on these aspects to reduce the time required for others to understand and verify an artifact.
%\end{itemize}
}

\framebox{
  \parbox{\linewidth}{
  Intel and Xeon are trademarks of Intel Corporation in the U.S. and/or other
  countries.

  \vspace{0.5em}
  Benchmark results were obtained prior to implementation of recent software
  patches and firmware updates intended to address exploits referred to as
  ``Spectre'' and ``Meltdown''. Implementation of these updates may make
  these results inapplicable to your device or system.

  \vspace{0.5em}
  Software and workloads used in performance tests may have been optimized for
  performance only on Intel\regtm{} microprocessors. Performance tests, such as
  SYSmark\othertm{} and MobileMark\othertm{}, are measured using specific
  computer systems, components, software, operations and functions. Any
  change to any of those factors may cause the results to vary. You should
  consult other information and performance tests to assist you in fully
  evaluating your contemplated purchases, including the performance of that
  product when combined with other products.

  \vspace{0.5em}
  For more information go to \url{http://www.intel.com/benchmarks}.
  }
}
~\\
\scriptsize\noindent\othertm{}Other names and brands may be claimed as the
property of others.